No effects of androgen receptor gene CAG and GGC repeat polymorphisms

on digit ratio (2D:4D): Meta-analysis

Martin Voracek *

*Department of Basic Psychological Research and Research Methods, School of Psychology,*

*University of Vienna, Austria*

*Georg Elias Müller Department of Psychology, Georg August University of Göttingen, Germany*

First ms. draft: Sep 22, 2012

This ms. version: June 14, 2013

* Corresponding author. Tel. +43-1-4277-47126; fax: +43-1-4277-47192.

*E-mail*: martin.voracek@univie.ac.at.

**ABSTRACT**

*Objectives*: A series of meta-analyses assessed whether differentially efficacious variants (CAG and GGC repeat-length polymorphisms) of the human androgen receptor gene are associated with digit ratio (2D:4D), a widely investigated putative pointer to prenatal androgen action.

*Methods*: Extensive literature search strategies identified a maximum of 16 samples (total $N$ = 2157) eligible for meta-analysis.

*Results*: In contrast to a small-sample ($N$ = 50) initial report, widely cited affirmatively in the literature, meta-analysis of the entire retrievable evidence base did not support associations between androgen receptor gene efficacy and 2D:4D.

*Conclusions*: These meta-analytical nil findings, along with several further suggestive strands of evidence consistent with these, undermine one validity claim for 2D:4D as a retrospective pointer to prenatal testosterone action.

**INTRODUCTION**

With several hundreds studies published over the past decade, the second-to-fourth digit ratio (2D:4D) presently is the most intensely investigated candidate pointer to prenatal androgen action (Hampson and Sankar 2012). 2D:4D is lower in men than women, sex and individual differences therein originate early during fetal development, and seem sufficiently stable thereafter. Hence, many believe 2D:4D appropriately indexes prenatal androgen action.

There is abundant evidence for long-term, permanent (i.e., organizing) effects of prenatal androgens on the brain, behavior, and disease susceptibility (Breedlove 2010). However, animal models of organizing sex-hormonal effects may not generalize unrestrictedly to humans, human fetal measurement of hormones is hazardous, human experimentation infeasible, and experiments of nature (endocrine disorders) have intrinsic limits of insight. It is therefore important to have simple, non-invasive retrospective indicators of prenatal androgen action. However, most validity cues for 2D:4D stem from correlational designs or animal experimentation. Evidence that genetically based differential efficacy of the human androgen receptor (AR) is correlated with 2D:4D (Manning et al. 2003) has been mentioned as the "strongest evidence that androgens affect digit ratio" (Breedlove 2010, p. 4117), accompanied by several further validity cues for 2D:4D discussed in the same review paper.

Explanatory background of this appraisal is as follows (Nieschlag and Behre 2004): the AR, an evolutionary well-conserved member of the nuclear receptor superfamily of ligand-activated transcription factors, is the core regulatory protein for androgen responsivity. The human AR gene, located in the pericentromeric region of the long arm of the X chromosome (Xq11.2-12), is encoded by eight exons. Exon 1 codes for the modulating N-terminal region, the first out of three functional domains of the AR protein. Genetic polymorphisms in the form of repetitive trinucleotide sequences within exon 1 code for two amino acid tracts, namely CAG (polyglutamine) and GGC (also termed GGN, polyglycine) stretches. Repeat-lengths of these two expansion polymorphisms vary interindividually and between populations, but not between the sexes and, of importance, are functional. That is, longer (vs. shorter) CAG (as well as GGC) repeats constitute loss-of-function (vs. gain-of-function) mutations, leading to less (vs. more) AR transcriptional activity (corresponding to the CAG polymorphism; see Buchanan et al. 2004; Chamberlain et al. 1994; Kazemi-Esfarjani et al. 1995; Tut et al. 1997; Wang et al. 2004) and smaller (vs. larger) amounts of AR protein (corresponding to the GGC polymorphism; see Brockschmidt et al. 2007; Ding et al. 2005; Lundin et al. 2007; Werner et al. 2006). More precisely, within the normal range of variation, research evidence fairly consistently suggests that these effects are linear, at least so for the CAG polymorphism (for an exception, see Gao et al. 1996). Exceeding critical thresholds, overly long CAG repeats lead to neurodegenerative disease, overly short CAG repeats predispose to prostatic neoplasia, whilst GGC repeat-length does not appear clearly disease-associated (Nieschlag and Behre 2004). In this latter context, it should be noted that the evidence concerning the linearity of the effect and its direction is inconsistent for the (less polymorphic; see Tut et al. 2007) GGC polymorphism. Whereas some studies support an inverse relationship between GGC repeat-length and AR transcriptional activity (Ding et al. 2005), others report a direct relationship, i.e., a directionally opposite effect (Brockschmidt et al. 2007; Werner et al. 2006), and still others report evidence for nonlinear effects (Lundin et al. 2007). These peculiarities or uncertainties have consequences for analyzing associations of GGC repeat-number with other variables, as they would call for detailed distributional, instead of merely correlational (see Zhang et al. 2013), data analysis. This point is taken up in the Discussion again.

Given that testosterone effects on 2D:4D would likely be mediated by the androgen receptor, a correspondence of longer (less efficacious) CAG stretches with higher (less masculinized) 2D:4D, i.e., a positive correlation between these variables, is expected. This has indeed been found in a first, small-sample study of CAG repeats and 2D:4D (Manning et al. 2003). The report is one of the most-cited of 2D:4D research (about 200 citations, according to Google Scholar) and its finding a validity argument for 2D:4D.

However, several more recent CAG/2D:4D studies evidently failed to replicate these initial findings (e.g., Hampson and Sankar 2012). Also, the generally much less thoroughly investigated GGC polymorphism has meanwhile been probed vis-à-vis 2D:4D (e.g., Zhang et al. 2013; to no avail). Accordingly, to reach clarification and reconciliation of what apparently has come to be a research literature with inconsistent findings, a series of meta-analyses of the entire retrievable empirical evidence on CAG/GGC repeats and 2D:4D was performed.

**METHODS**

Studies eligible for the meta-analyses (investigating CAG/GGC repeats with right-hand, left-hand, and right-minus-left-hand digit ratios; henceforth R2D:4D, L2D:4D, and $\Delta_{R-L}$) were identified via four integrated literature search strategies up to May 2013: databases (Web of Science, Scopus, PubMed, PsycInfo, Google Scholar, WorldCat, UMI/ProQuest Dissertations), including extensive online searches in the latter three databases for unpublished reports; cited reference searches for Manning et al. (2003); a research bibliography (Voracek and Loibl 2009); and personal contacts with researchers.

These combined efforts yielded 11 published studies (including one published journal abstract: Mas et al. 2009), providing a maximum of 16 samples (total $N$ = 2157) which originate from 8 countries (Australia, Canada, China, Slovakia, Spain, Tanzania, UK, USA). Study details and findings are displayed in Table I. Unpublished or non-English accounts could not be located. One study (Latourelle et al. 2008) cannot be found via usual searches in databases, because it is not properly indexed with expected search terms. Instead, this report was identified via a research bibliography (Voracek and Loibl 2009). Continued requests to study authors for unreported, additional, or updated study details (not fulfilled in all cases) ensued in early 2012 and were followed up until May 2013. For additional citation analyses, studies' citation counts (as of May 2013) from Google Scholar were utilized (adjusted for time since publication, i.e., cites/yr.). Although this implicit assumption of linear increase in citations may be questioned, cumulative citation analysis of the most-cited report (Manning et al. 2003) in Web of Science showed that its citations indeed accrued practically linear ($r$ = .99 over the years 2003-2012). It is emphasized that not the absolute citation numbers were used for the citation analyses, but rather the papers' citation density per time unit (cites/yr.; see Table I). This adjusts for the differential time elapsed since publication and thus enables meaningful comparisons between more (vs. less) recently published reports.

Following meta-analytical standard practice (Lipsey and Wilson 2001, p. 70), unreported non-significant effects in the original studies, with additional results details not received in personal communication, were set to zero, whereas effects reported as significant, but lacking numerical detail, to just-significant (i.e., $r$ corresponding to $p = .05$, two-tailed). For female samples, $r$ for the biallelic mean (i.e., average CAG/GGC number of the two alleles), as reported in all respective primary studies, was used. This method, although common in AR research (Rajender et al. 2007, p. 167), is less than optimal and complicates the interpretation of findings for women. This is because the AR gene is located on the X chromosome. Hence, two AR alleles are present in women, of which one is randomly inactivated (Migeon 2007). All primary studies with female samples retrieved for the meta-analysis used the biallelic mean for analysis (see Table I). One recent study (Zhang et al. 2013) in addition reported separate analyses (all not significant as well) for the shorter vs. the longer allele among women. Some studies of the broader AR literature mention analyses by taking only the actually active AR allele into consideration, which method would be superior (Migeon 2007; Rajender et al. 2007, p. 167). However, no such 2D:4D studies of AR variation among women could be found. For one study (Butovskaya et al. 2012), sample amplification since publication resulted in a loss of the sole significant finding previously reported (for L2D:4D; the updated effect size was used for analysis). In the course of one longitudinal study (Knickmeyer et al. 2011), digit ratios were measured thrice (2 weeks post term, and at 1 yr. and 2 yrs. of age), with some study attrition for the follow-up measurements. Hence, meta-analytically combined mean $r$ values over these three waves of data collection and the mean $N$ thereof were taken as conservative estimates for analysis (see Table I for the details of the above decisions made for the meta-analyses). Comprehensive Meta-Analysis version 2.2 software was used, and, following standard practice for combining correlation coefficients, Fisher's $z$ transformation was applied.

**RESULTS**

The meta-analytical findings are summarized in Table I. Theory-compliant correlations of R2D:4D, L2D:4D, and $\Delta_{R-L}$ with CAG or GGC repeats all would be positive. As for the CAG studies, fixed-effect modelling (assuming between-study effect heterogeneity not exceeding sampling error, i.e., chance) indicated that effects on digit ratios were essentially nil. Tests for effect heterogeneity ($Q$ statistics) never were significant and heterogeneity effect sizes ($I^2$) never substantial (when $I^2 > 75\%$); hence, random-effects modelling (appropriate for substantial between-study effect heterogeneity beyond sampling error) was not necessary.

In line with these central findings, subgroup analyses, testing for conceivable effect moderators, namely age group (adult vs. non-adult samples), finger-length measurement method (direct vs. image-based), and participant sex (male vs. female), were not suggestive of consistent effect moderation (details of these and further supplemental analyses, below, omitted for brevity). Cumulative meta-analyses (studies added chronologically one by one to the model) showed that the overall effect lost significance with either adding the very next study (for R2D:4D: Latourelle et al. 2008) or later on (for $\Delta_{R-L}$: with Loehlin et al. 2012 added) and never gained back significance. The effect for L2D:4D never was significant at any time. Sensitivity analyses (one study removed at a time from the model) did not indicate any influential studies in the model, meaning that overall meta-analytical results did not critically depend on any of the assumptions made for analysis (see Methods). By the same token, sensitivity analyses also ruled out an influence of several mixed-ethnic samples on overall results (90%, 87%, 80%, and 56% Caucasians in the samples of Hampson and Sankar 2012; Loehlin et al. 2012; Knickmeyer et al. 2011; and Hurd et al. 2011; Caucasians-only subsample analyses, as partly reported in these studies, left the respective total-sample results unchanged). Visual inspection of funnel plots, as well as several formal statistical tests for publication bias (Begg-Mazumdar rank-correlation test, Egger regression-intercept test, and Duval-Tweedie trim-and-fill method), did not yield formal signs for publication bias in the literature. However, power of these tests is known to be low in smaller meta-analyses.

Meta-regression models, with study effects regressed on publication year, paralleled the evidence from the cumulative meta-analyses, in that more recent studies had smaller effects (for $\Delta_{R-L}$: significant negative regression slope over time, $p$ = .008). Additional meta-regressions, with study cites/yr. regressed on study effects, evidenced citation bias for this research literature, i.e., studies with larger effects were cited more frequently within the same time unit (for $\Delta_{R-L}$: significant positive regression slope with cites/yr., $p$ = .004).

As for the much smaller set of GGC studies, fixed-effect modelling was appropriate as well. GGC numbers were weakly positively (albeit nominally significantly) related only to R2D:4D (combined $r$ = .080, 95% $CI$ = .011-.148, $p$ = .023). However, sensitivity analysis showed that this effect was crucially dependent exactly on the one sample from Mas et al. (2009) for which the effect magnitude had to be estimated. Without this sample, significance would be lost. Hence, not much credence should be given to this modest effect, arising from a limited evidence base. Owing to the paucity of GGC studies, none of the above supplemental analyses for CAG studies were feasible for GGC studies.

**DISCUSSION**

Meta-analysis of the accumulated retrievable evidence on effects of functional AR gene variants on 2D:4D does not support initial evidence (Manning et al. 2003) for such associations. The empirical base for this conclusion meanwhile is diverse, comprising 11 reports without any author overlap (i.e., truly independent replication attempts), from 8 countries located on 5 continents, featuring major ethnic groups (Africans, Caucasians, and East Asians), including an indigenous hunter-gatherer group (the Hadza people), different 2D:4D measurement methods (direct vs. various image-based), women and male-to-female transsexuals (in addition to men), and non-adult samples from various developmental stages (pubertal individuals and adolescents, newborn and toddlers) in addition to adults.

Despite this noticeable diversity of study features, all studies subsequent to Manning et al. (2003) are akin in that the initially published magnitude of the effect for R2D:4D and $\Delta_{R-L}$ failed to replicate by far. In this context of reproducibility of research findings and appraisal of the cumulative evidence through meta-analyses, it has been noted that "multiple replication attempt effect sizes that are homogenous around zero (without reasons for the original effect to differ) suggest that the original effect was a fluke" (Ijzerman et al. 2013, p. 128). In fact, there is only one nominally significant effect later on in this literature (Loehlin et al. 2012; for L2D:4D among women), and meta-analytical aggregation of the entire literature demonstrates that the effects overall are nil.

Technically, meta-analysis is feasible with as few as two studies. Also, it already is meaningful and justified in such a situation, because even a two-study collection of related evidence will have higher statistical power than either one of those studies (Valentine et al. 2010). Hence, although the current evidence base for the meta-analysis of GGC variation and 2D:4D admittedly is small, there is a gain of information from quantitatively combining the retrievable 4 samples (Table I). Mainly for technical challenges concerning amplification for analysis (Rajender et al. 2007, p. 169), the GGC polymorphisms has been markedly less commonly studied than the CAG polymorphism. The relative size of these literatures is reflected in the different numbers of samples available for the meta-analyses (Table I). Consequently, very few studies have investigated effects of CAG and GGC polymorphism combined, and no such joint distributional analysis of these two polymorphisms is currently available vis-à-vis 2D:4D. Such studies would be all the more interesting because of two points: first, there is evidence for linkage equilibrium, i.e., lack of association, between these two polymorphisms (Ding et al. 2005; Kittles et al. 2001), despite the close proximity of the two gene loci (separated by 248 amino acids of non-polymorphic sequence or only by 1.1 kb; see Kittles et al. 2001; Rajender et al. 2007). And second, comparisons of the physiological effects of the (frequently studied) CAG polymorphism and the (much less frequently studied) GGC polymorphism suggest that GGC repeat-number effects might be larger than those of CAG (Ding et al. 2005). Further, as already discussed in the Introduction, the inconsistent research evidence for the direction and linearity of GGC repeat-number effects would call for detailed distributional analysis, but no such paper is yet available in the 2D:4D literature. An additional point of consideration, concerning the CAG polymorphism, is that this repeat has a different location within the N-terminal region in the human AR (located upstream of the activation domain), as compared to rodent (mouse and rat) AR receptors (located within the activation domain), which location differences of the tract relative to the activation domain may play a role in observed differences in the activity of human vs. rodent AR receptors (Chamberlain et al 1994). These differences would make cross-species extensions of the human CAG/2D:4D literature to rodents valuable. However, similar to the above-mentioned points, as of yet no such studies have been conducted.

Even with small numbers of primary studies eligible for meta-analysis, this technique remains the most powerful, reliable, valid, exhaustive, and unbiased method of appraising and synthesizing empirical research evidence, because all alternatives are less transparent and defensible and thus less reproducible and valid (Valentine et al. 2010). For instance, the narrative conclusions of traditional reviews are derived from unique "cognitive algebra" synthesis of reviewers, with idiosyncratic and unstated rules applied. Conversely, in the meta-analytical framework any undue emphasis on the results of individual studies is avoided. As for another example, the vote-counting method (i.e., synthesizing studies, using the outcomes of individual statistical tests) inevitably confounds magnitude of effects with sample size and, perhaps counterintuitively, actually has less statistical power, the more evidence accumulates (Valentine et al. 2010). In contrast, the gain in power and thus reduction of uncertainty due to meta-analysis can be seen in comparing the length of the confidence interval for the combined effect of CAG variation on R2D:4D from 16 samples (95% *CI* = -.021 to .064; see Table I) with the corresponding confidence interval from the Manning et al. (2003) study alone (95% *CI* = .013 to .526): the former one is six times shorter than the latter one.

As for the typical size of meta-analyses from comparable research fields, a systematic survey of 49 meta-analyses published 2011 in the field of neuroscience (Button et al. 2013) yielded a median number of 12 primary studies (interquartile range: 7 to 17 studies) per meta-analysis. Another study (Davey et al., 2011) analyzed the entire collection of the Cochrane Database of Systematic Reviews, which is the major resource of systematic reviews on the effects of healthcare intervention. Of a total of 22453 meta-analyses found in 2321 Cochrane reviews, 36% of all meta-analyses included just the minimum of 2 primary studies, the median number was 3, and the interquartile range 2 to 6 primary studies. By these standards, the current meta-analysis on CAG variation and 2D:4D is not small. Instead, sufficient evidence appears now to have been accumulated to draw conclusions about the veridicality of the effect suggested by Manning et al. (2003).

Publication patterns in the literature meta-analyzed here are such that a spate of replication attempts appeared only quite recently (2011-2013). The unreplicated first report continues to be the most frequently cited in this line of inquiry, even with adjustment for time since publication (i.e., cites/yr.). There appears to be citation bias for this literature, as suggested by a positive meta-regression slope for study cites/yr. regressed on study effect size. Citation bias is a phenomenon analogous to publication bias, albeit operating more subtle, and is a known pervasive problem in biomedical and other empirical research (Greenberg 2009; Jannot et al. 2013). Apparently rooted in cognitive biases (foremost, confirmation bias and attentional bias), citation bias distorts the perceptions of the available scientific evidence among users of scientific literature.

Apart from these general observations, which seem to apply for the literature meta-analyzed here as well, it is informative to look further into the details of why and in which ways the Manning et al. (2003) report continues to be cited by subsequent research in this field. To address these questions, the most recent citations it received were scrutinized. This additional qualitative, context-specific, and content-analytic citation analysis (see Greenberg 2009) was also based on Google Scholar. In May 2013, a total of 44 citations were ascertained which the Manning et al. (2003) report received by papers published, archived, or appearing online ahead of print in the years 2012 and 2013. The onset of this observation period was selected to be after the first failed replications of Manning et al. (2003) were published (see Table I). The following results from this qualitative citation analysis appear noteworthy: first, 27 of 44 (61%) papers citing Manning et al. (2003) cite it solely, i.e., no other papers from the CAG/2D:4D literature are cited. Second, 35 of 44 (80%) citations are confirmative, i.e., are brought forward as evidence for 2D:4D and CAG repeat number being correlated. Third, 4 of 44 (9%) of the papers citing Manning et al. (2003) are part of the (more recent) literature meta-analyzed here, and all of these replication attempts (4 of 4; 100%) cite at least one further paper on 2D:4D and CAG repeat number. However, only a minority of the remainder (13 of 40; 33%) of papers from 2012 or 2013 citing Manning et al. (2003) cite at least one of the papers that failed to replicate it. Fourth, related to this, some evident replication failures (Hurd et al. 2011; Knickmeyer et al. 2011) are sometimes cited confirmatively instead (i.e., as "suggestive"), on the basis of highlighting selected findings within one study or calculating one-tailed $p$ values post hoc. Also, mimicking the vote-counting method, citations of Manning et al. (2003) sometimes are contrasted ("but see") with accompanying citations of one or two of its failed replication studies, as if the respective research evidence were divided, mixed, or uncertain, which is in stark contrast to both the actual size of the literature and the clear meta-analytical null findings for this literature. Fifth, all of the most recent citations (44 of 44; 100%) of Manning et al. (2003) were from within the 2D:4D literature, i.e., the paper appears not to have been cited in other, related research contexts during this recent observation period. Taken together, these observations from the additional qualitative citation analysis of Manning et al. (2003) make a strong case for the necessity of this meta-analysis, because the initial, unreplicated report continues to be cited frequently, very often confirmatively, and oftentimes solely, whilst its non-replications are cited much less often, sometimes are reinterpreted, and the overall evidence cannot be accurately ascertained without meta-analysis.

Manning et al. (2003) was the second-smallest sample included in the meta-analysis, whereas the latest study (Zhang et al. 2013) had a total sample size almost 14 times larger and the meta-analysis altogether one more than 40 times larger than the first study. Due to the small sample ($N$ = 50) of Manning et al. (2003), there was considerable uncertainty about the actual magnitude range of the observed effects, as indicated through wide 95% confidence intervals (.013-.526 and .091-.580) for the study effects ($r$ = .29 and .36 for R2D:4D and $\Delta_{R-L}$, respectively). Accordingly, power analysis suggests that this first study was underpowered, as a sample of $N$ = 50 merely has 41% power to detect medium-sized population correlations ($\rho$ = .25) with two-tailed testing. In contrast, the corresponding power figure for the male sample ($N$ = 294) in the so far largest study (Zhang et al. 2013) is 99% and thus optimal.

The above additional calculations (confidence intervals and power) for the Manning et al. (2003) study, in tandem with the findings of the meta-analysis, illustrate some important points about the associations of study power with the reproducibility of study findings, which are relevant for other subliteratures of 2D:4D research as well (see Voracek 2013a, 2013b). Underpowered studies are an endemic problem in empirical research (Button et al. 2013). Although it is well-known that low power reduces the chances of detecting true effects, it is less recognized that low power also makes it less likely that published statistically significant study results actually reflect a true effect. Given common publication practices, in this latter situation the positive predictive value of a claimed effect is low and, should the effect be true, its magnitude likely is overestimated, thus leading to generally low reproducibility of results (Button et al. 2013). The meta-analytical results suggest that this is the case with the AR gene literature of 2D:4D research.

Apart from the current meta-analytical null findings, six additional strands of evidence are not suggestive of major AR gene-related effects (by implication, X-chromosomal in humans) on 2D:4D. The following qualitative synthesis of evidence partly draws on conclusions previously made in separate accounts (e.g., Berenbaum et al. 2009; Hampson and Sankar 2012; Loehlin et al. 2012; Voracek and Dressler 2009), but updates these with more recent findings and assembles them in their entirety here for the first time. First, no androgen-related gene locations (or in linkage disequilibrium with these), including the AR gene, or any other X-chromosomal locations, have been implicated in the formation of 2D:4D in so far two large-scale genome-wide association studies (partly featuring discovery and additional independent replication samples: Medland et al. 2010; Lawrance-Owen et al. 2013). Second, in this context, it is interesting to note that one CAG study included in the meta-analysis (Loehlin et al. 2012) pursued a twins-and-their-siblings design. This enables controls for many shared genetic, parental, and prenatal factors otherwise uncontrolled in conventional samples. Of note, these informative within-sibship tests of Loehlin et al. (2012) similarly yielded null results: siblings' CAG-repeat difference did not predict their difference in 2D:4D. Third, the AR gene has also not been implicated for 2D:4D formation in zebra finches (a species in which the AR gene is not located on a sex chromosome; Forstmeier et al. 2010).

Fourth, so far three human family studies of 2D:4D (Ramesh and Murty 1977; Voracek and Dressler 2009; Ventura et al. 2013) yielded no evidence of family correlation patterns indicative of X-linked inheritance ($r_{\text{sister-sister}} > r_{\text{brother-brother}} > r_{\text{brother-sister}}$ and $r_{\text{father-daughter}} = r_{\text{mother-son}} > r_{\text{mother-daughter}} > r_{\text{father-son}}$). Fifth, another sign of X-linked inheritance is greater male than female intrasex trait variability. However, no such sex difference in the variance of 2D:4D has been demonstrated.

Sixth, there is further evidence from a first study of individuals with complete androgen-insensitivity syndrome (CAIS; Berenbaum et al. 2009; for discussions, see Breedlove 2010; Forstmeier et al. 2010). Genetically, CAIS individuals are male (46,XY), but phenotypically female, because of completely non-functional AR. In the study, CAIS individuals' 2D:4D was indistinguishable of normal women's 2D:4D (who have normal, i.e., differentially functional, AR), and thus, relative to male controls, it was not dramatically feminized (for the latter group comparison of CAIS and male controls, effects were $r = .24$ and $.14$ for R2D:4D and L2D:4D; calculated from Berenbaum et al. 2009). Hence, even the complete AR inactivation in CAIS individuals merely leads to small-to-medium 2D:4D deviations, relative to normal-AR men. Furthermore, despite being all identical with respect to complete AR inactivation, CAIS individuals did not have reduced variability in 2D:4D, but rather showed sample variation therein just as much as normal male or female controls. Hence, the entirely preserved individual differences in their 2D:4D must be due to other factors than testosterone action. Limitations of this evidence of Berenbaum et al. (2009) include the very small case-group size (16 CAIS individuals) and the reliance on an unmatched case-control design, instead of utilizing matched or nested case-control comparisons. Hence, replications would be worthwhile.

In conclusion, this meta-analysis suggests that AR efficacy (via CAG/GGC repeats) is unrelated to 2D:4D, at least within the normal range of repeats. Six further lines of evidence and arguments assembled here are consistent with this conclusion. Taken together, this undermines one validity claim for 2D:4D as a prenatal-androgen pointer (for a discussion and evaluation of other validity cues for 2D:4D, see Breedlove 2010). If efficacy of the AR, the core regulator of androgen action, has no effect on 2D:4D, it is difficult to understand in what other equally important ways testosterone could act on 2D:4D formation, such that 2D:4D could reliably reflect testosterone action. Although masculinizing effects of testosterone through aromatization to estradiol and subsequent estrogen-receptor binding have some role in rodent species, this route does not appear to play a dominant role in masculinization in humans or other primate species (see Hampson and Sankar 2012). Also, estrogens feminize digit ratio in mice (Zheng and Cohn 2011), which would argue against the aromatization pathway of masculinization for rodent 2D:4D specifically.

Could 2D:4D still reflect prenatal androgens, even if not correlated with functional AR gene variation? In this scenario, inhibitory feedback influences would effectuate a correspondence of less sensitive AR to increased serum androgen levels and vice versa. These two factors could then cancel each other out. A premise of this idea would be evidence for positive correlations of CAG repeats and circulating testosterone. It however appears that this assumption is not well-supported empirically. Among women, Westberg et al. (2001) found a negative, instead of positive, correlation, whereas Skrgatic et al. (2012) none. In the latter study, a positive correlation was only found among a patient group of women affected with PCOS, and only for total testosterone, not for free testosterone. Among boys, Durdiaková et al. (2013) found no correlation, whereas Mouritsen et al. (2013) observed a complex effect (nonlinear association) at one measurement point, but no such (or other) effects for two further measurement points within the same cohort.

Rather than recommending "further studies are necessary" (Zhang et al. 2013, p. 105), the current meta-analytical null results suggest that reconsideration and rethinking of assumptions in this line of inquiry and pursuing alternative approaches for elucidating 2D:4D formation, such as genome-wide association studies, might be opportune.

Table I. Correlations of 2D:4D with androgen receptor gene CAG and GGC polymorphisms: individual studies and meta-analysis.

| Study | Cites/yr. | Country | Sample | 2D:4D measurement | N | r | | |
|---|---|---|---|---|---|---|---|---|
| | | | | | | R2D:4D | L2D:4D | $\Delta_{R-L}$ |
| *CAG studies* | | | | | | | | |
| Manning et al. (2003) | 18.0 | UK | men (sports club members, university staff and students) | direct | 50 | .29* | .005 | .36** |
| Latourelle et al. (2008) | 0.8 | USA | men | photocopies | 35 | .00[a] | – | – |
| Latourelle et al. (2008) | 0.8 | USA | women | photocopies | 72 | .00[a] | – | – |
| Mas et al. (2009) | 0.0 | Spain | men | photocopies | 72[b] | -.0685[b] | -.054[b] | .002[b] |
| Mas et al. (2009) | 1.0 | Spain | male-to-female transsexuals | photocopies | 63[b] | .0021[b] | -.0941[b] | .1447[b] |
| Hurd et al. (2011) | 9.7 | Canada | men | digicam photographs | 178-180 | .006 | -.12 | .14 |
| Knickmeyer et al. (2011) | 6.0 | USA | male newborns and toddlers | photocopies | 71-74[b] | .143[b] | .014[b] | .108[b] |
| Knickmeyer et al. (2011) | 6.0 | USA | female newborns and toddlers | photocopies | 70-74[b] | -.133[b,c] | -.028[b,c] | -.010[b,c] |
| Butovskaya et al. (2012) | 2.5 | Tanzania | men (Hadza hunter-gatherers) | direct | 103[b] | .1347[b] | .1913[b] | -.0798[b] |
| Folland et al. (2012) | 1.0 | UK | men (young adults) | photocopies | 71 | .10 | .20 | .00[a] |
| Hampson and Sankar (2012) | 2.0 | Canada | men (students) | flatbed scans | 134 | -.085 | -.063 | -.047 |
| Loehlin et al. (2012) | 2.0 | Australia | male adolescents | photocopies | 182 | -.06 | -.13 | .10 |
| Loehlin et al. (2012) | 2.0 | Australia | female adolescents | photocopies | 218 | .08[c] | .14*[c] | -.06[c] |
| Durdiaková et al. (2013) | 0.0 | Slovakia | boys and male adolescents | flatbed scans | 147 | .04 | .09 | -.085[b] |
| Zhang et al. (2013) | 1.0 | China | men (students) | photocopies | 294 | .003 | .016 | -.022 |
| Zhang et al. (2013) | 1.0 | China | women (students) | photocopies | 391 | .030[c] | -.018[c] | .055[c] |
| Samples (total *N*) | | | | | | 16 (2157) | 14 (2051) | 14 (2044) |

| Study | Cites/yr. | Country | Sample | 2D:4D measurement | $N$ | $r$ | | |
|---|---|---|---|---|---|---|---|---|
| | | | | | | R2D:4D | L2D:4D | $\Delta_{R-L}$ |
| Combined $r$ [95% $CI$] | | | | | | .021 [-.021 to .064] | .006 [-.038 to .049] | .026 [-.018 to .069] |
| $Q$ ($I^2$) | | | | | | 12.3 (0%) | 19.1 (32%) | 16.8 (22%) |
| *GGC studies* | | | | | | | | |
| Mas et al. (2009) | 1.0 | Spain | men | photocopies | 72[b] | .196*[d] | .196*[d] | .196*[d] |
| Mas et al. (2009) | 1.0 | Spain | male-to-female transsexuals | photocopies | 63[b] | .000[a] | .000[a] | .000[a] |
| Zhang et al. (2013) | 1.0 | China | men (students) | photocopies | 294 | .095 | .106 | -.026 |
| Zhang et al. (2013) | 1.0 | China | women (students) | photocopies | 391 | .060[c] | .023[c] | .044[c] |
| Samples (total $N$) | | | | | | 4 (820) | 4 (820) | 4 (820) |
| Combined $r$ [95% $CI$] | | | | | | .080* [.011 to .148] | .066 [-.003 to .134] | .029 [-.040 to .097] |
| $Q$ ($I^2$) | | | | | | 1.6 (0%) | 2.7 (0%) | 3.0 (8%) |

[a]Unreported in original study, additional results details not received (hence, effect set to zero). [b]Unreported in original study or sample amplified since then (supply of additional results details gratefully acknowledged). [c]Correlation for biallelic mean. [d]Reported as significant in original study, additional results details not received (hence, effect set to just-significant, $p$ = .05, two-tailed).

*$p$ < .05, **$p$ < . 01 (two-tailed).